\documentclass[aps,reprint,superscriptaddress]{revtex4-1}
\usepackage{amsmath}
\usepackage{amsfonts}
\usepackage{dsfont}
\usepackage{subfigure}
\usepackage{graphicx}
\usepackage{float}
\usepackage{bm}
\usepackage{hyperref}

\begin{document}


\title{Coherence as resource in scattering quantum walk search on complete graph}



\author{Yun-Long Su}
\affiliation{Institute of Modern Physics, Northwest University, Xi'an 710069, China}
\affiliation{School of Physics, Northwest University, Xi'an 710069, China}
\affiliation{Shaanxi Key Laboratory for Theoretical Physics Frontiers, Xi'an 710069, China}
\author{Si-Yuan Liu}\email{syliu@iphy.ac.cn}
\affiliation{Institute of Modern Physics, Northwest University, Xi'an 710069, China}
\affiliation{Shaanxi Key Laboratory for Theoretical Physics Frontiers, Xi'an 710069, China}
\author{Xiao-Hui Wang}
\affiliation{School of Physics, Northwest University, Xi'an 710069, China}
\affiliation{Shaanxi Key Laboratory for Theoretical Physics Frontiers, Xi'an 710069, China}
\author{Heng Fan}
\affiliation{Institute of Modern Physics, Northwest University, Xi'an 710069, China}
\affiliation{Shaanxi Key Laboratory for Theoretical Physics Frontiers, Xi'an 710069, China}
\affiliation{Institute of Physics, Chinese Academy of Sciences, Beijing 100190, China}
\author{Wen-Li Yang}
\affiliation{Institute of Modern Physics, Northwest University, Xi'an 710069, China}
\affiliation{Shaanxi Key Laboratory for Theoretical Physics Frontiers, Xi'an 710069, China}



\date{\today}

\begin{abstract}
We investigate the behavior of coherence in scattering quantum walk search on complete graph
under the condition that the total number of vertices of the graph is greatly larger than the marked
number of vertices we are searching, $N \gg v$.
We find that the consumption of coherence represents the increase of the success probability for the searching,
also the consumption of coherence is related to the efficiency of the algorithm represented by oracle queries.
If no coherence is consumed, the efficiency of the algorithm will be the same as the classical blind search, implying that
coherence is responsible for the speed up in this quantum algorithm over its classical counterpart.
In case the initial state is incoherent, still $N \gg v$ is assumed,
the probability of success for searching will not change with time, indicating that this quantum search algorithm loses its power.
We then conclude that the coherence plays an essential role and is responsible for the speed up in this quantum algorithm.
\end{abstract}


\maketitle

\section{INTRODUCTION}
Random walk is an important prototype for efficient classical algorithms \cite{motwani1995,schoning1999probabilistic}.
As the quantum analogy of random walk, quantum walk \cite{Quantum_random_walks} is important in developing efficient quantum algorithms.
As we known, the quantum algorithms may have great speed up compared with their classical counterparts as shown by
Shor's algorithm and Grover search \cite{shor1999polynomial,grover1997quantum}.
There are two kinds of quantum walk, discrete quantum walk and continuous quantum walk. Both of them show their advantages over the classical random walks \cite{ambainis2001one,childs2003exponential}.
A quantum search algorithm constructed from discrete quantum walk on hypercube is shown that
it has the similar boost over the classical search algorithms as the Grover search algorithm \cite{shenvi2003quantum}.
A search algorithm based on continuous quantum walk also has quadratic speedup  \cite{childs2004spatial,childs2004spatial}.
Quantum walk is not only studied in theory but also realized in experiment via various ways \cite{schmitz2009quantum,xue2009quantum,peruzzo2010quantum,zahringer2010realization,xue2014trapping,bian2015realization,xue2015experimental}.
To describe the implementation of quantum walk using linear optical elements, scattering quantum walk is proposed \cite{Scattering_quantum_walk}.
An algorithm to search based on such quantum walk is constructed and analyzed on complete graph explicitly, showing quadratic speed up as well \cite{reitzner2009quantum}.

On the other hand, we know that quantum entanglement plays a significant role in
quantum teleportation, super-dense coding and quantum phase transitions in many-body systems,
\cite{horodecki2009quantum,osterloh2002scaling}. Also other quantum correlations, such as quantum discord, are
critical in quantum information and many-body systems \cite{zurek2000einselection,ollivier2001quantum}. All of them can be assumed to be some kinds of resources for quantum information processing \cite{osterloh2002scaling,osborne2002entanglement,pirandola2013quantum,streltsov2012quantum}.
Recently, the resource theory of quantum coherence is proposed, and is attracted much attention \cite{baumgratz2014quantifying,winter2016operational,chitambar2016assisted}.
Remarkably, coherence is shown to be valuable resources in several well-known quantum algorithms.
Explicitly, it is shown that coherence is a resource in Deutsch-Jozsa algorithm based on quantum walk \cite{hillery2016coherence}.
Also, coherence depletion is shown to be related with probability of success in Grover search algorithm \cite{shi2016coherence}.
Coherence is assumed as resource in deterministic quantum computation with one qubit \cite{matera2016coherent}.
The interference in quantum walk makes quantum walk differ from the classical random walk,
so it is expected that coherence should play a key role in quantum walk algorithms.
In particular, we may wonder whether quantum coherence is directly responsible for the speed up of the quantum walk search
based on complete graph studied in Ref.\cite{reitzner2009quantum}.

In this paper, the role of coherence is studied systemically in a quantum walk search algorithm, the scattering quantum walk search \cite{reitzner2009quantum}.
We consider the condition that the total number of vertices $N$ of the graph, which is the scope of the data,
is greatly larger than the marked
number of vertices $v$ we are searching, $N \gg v >1$. The quantum search algorithm demonstrates
great advantage over classical ones in this condition. Two measures of coherence defined in \cite{baumgratz2014quantifying} is used to give the dynamics of coherence
and relate the searching probability of success with coherence.
In the progress of the algorithm, the coherence is decreasing with the increasing of probability of success.
The coherence reaches its minimum when the success probability is maximal.
When reducing the efficiency of the algorithm, while the minimum of coherence increases,
the connection between the coherence and probability of success still exists.
Besides, when there is no coherence consumed, this quantum search algorithm will have the same complexity as that of the classical blind search algorithm.
If the consumption of coherence is below a proper value, the efficiency of the algorithm will be less efficient than classical search with memory.
Further more, when initial states are incoherent, also $N \gg v$,  the probability of finding the targets almost keeps unchanged
compared to that of the start stage.
So we cannot use such a state with no coherence to perform the search algorithm.
Based on those results, we conclude that the coherence plays an essential role
and is responsible for the speed up.

The paper is organized as follows: In Sec. \ref{sec:dynamic_of_coherence_in_quantum_walk_search},
we introduce the scattering quantum walk search and give the dynamics of coherence in two different measures.
The connection between the probability of success and coherence is presented.
In Sec. \ref{sec:reduce_efficiency},
we reduce the efficiency of the algorithm by choosing different phase shift and show numerically
that there is a connection between consumption of coherence and the efficiency of the algorithm.
In Sec. \ref{sec:walk_with_no_coherence}, we study the probability of success with an incoherent initial state.
Sec. \ref{sec:conclusion} is a summary of our results.

\section{Dynamic of coherence in quantum walk search} 
\label{sec:dynamic_of_coherence_in_quantum_walk_search}

The quantum walk search algorithm we studied here is the scattering quantum walk search \cite{reitzner2009quantum}. Scattering quantum walk \cite{Scattering_quantum_walk} is one of the versions of discrete quantum walk and is unitary equivalent to the coined quantum walk \cite{andrade2009equivalence}. It could be considered as the discrete quantum walk in optical network.

\begin{figure}[htbp]
 {
 \includegraphics[width=0.45\textwidth]{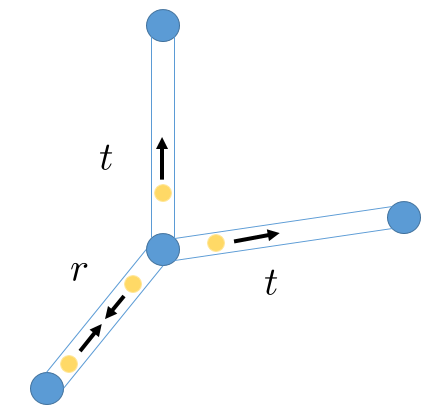}
 }
 \caption{The blue dots are vertices and yellow dots can be regarded as walkers. Together with the black arrow, the yellow dots represent the edge states. This figure can be understood as a photon traveling to a vertex with probability of $|r|^2$ being reflected back and $|t|^2$ being transfered to other vertices.}\label{SQW}
 \end{figure}

The scattering quantum walk is defined on a graph $\mathcal{G}(V,E)$ with $V$ being the set of the total vertices and $E$ being the set of edges connecting vertices. In this quantum algorithm, the particle walks on the edges of the graph rather than on the vertices. The Hilbert space in this algorithm is defined as
\begin{equation}
\mathcal{H}=l^2(\{|m,l\rangle \ m,l \in V,ml \in E  \}),
\end{equation}
where state $|m,l\rangle $ is an edge state going from vertex $m$ to $l$. The evolution of this quantum walk is defined by the local unitary operator for each vertex. Following the notation in \cite{reitzner2009quantum}, we denote $\Gamma(l)$ as the set of vertices connected to vertex $l$ and $\Gamma(l;k)$ as the set of vertices connected to vertex $l$ excluding vertex $k$. The local unitary operator for each vertex is defined as
\begin{equation}
U^{l} |k,l\rangle= -r^{l} |l,k\rangle + t^{l}\sum_{v \in \Gamma(l;k)} |l,v\rangle.
\end{equation}
Here $r^{l}$ and $t^{l}$ may be different for each vertex. This local operation transforms the state going into the vertex to the state scattering out of the vertex, illustrated in Fig. \ref{SQW}. For simplicity, we call the vertices we want to find as the marked ones and other vertices as the normal ones. Then, for normal vertices, we set
\begin{equation}
t=\frac{2}{|\Gamma(l)|},\ r=1-t
\end{equation}
and for marked ones
\begin{equation}
t=0, \ r=-e^{ i \varphi} ,
\end{equation}
where $|\Gamma(l)|$ is the number of vertices in the set $\Gamma(l)$. The local unitary operators for the normal one and marked one are denoted as $U^l_{0} $ and $U^l_{1}$ respectively.

In search algorithm, oracles are widely used to tell us if the element giving the query is the marked one. In this algorithm, the elements are the vertices of the graph. If we denote the set of marked vertices as $\mathcal{V}$ and the set of all vertices as $\mathcal{N}$, the oracle can be defined as a function of vertex
\begin{eqnarray}\label{oracle}
f(x)=
\begin{cases}\label{oracle}
&1 \quad x \in \mathcal{V} \\
&0 \quad x \in \mathcal{N} \backslash \mathcal{V}
\end{cases}\ .
\end{eqnarray}
In this quantum algorithm, a controlled unitary operator works as an oracle, i.e.
\begin{equation}
\mathcal{C}\hat{V}_{f}:\ |x\rangle \otimes |m\rangle \mapsto |x\rangle \otimes |m \oplus f(x)\rangle ,
\end{equation}
where the first one is a state of vertex and the second one is a qubit. When we run this algorithm, the oracle will send back the result according to Eq.(\ref{oracle}) and store it in the second qubit. Because $\mathcal{C}\hat{V}_{f}$ acts on $|x\rangle \otimes |m\rangle$, two extra states, a state of the vertex and a qubit, are necessary to be added to the state of walker for the use of oracle.
If we set the state of the walker as $|\psi_{n}\rangle$, the state of the whole system is the direct product of these states $|\psi_{n}\rangle \otimes |0\rangle \otimes |0\rangle $, before iterating the algorithm. To implement the search, we use a controlled unitary operator $\mathcal{C}\hat{W}_1$ which maps the state $|k,l\rangle \otimes |0\rangle \otimes |0\rangle $ to $|k,l\rangle \otimes |l\rangle \otimes |0\rangle$.
This step is required for the well function of oracle. Then the controlled unitary operator $\mathcal{C}\hat{V}_{f}$ will be applied to the state. When the result of the oracle is stored in the qubit, another controlled unitary operator $\mathcal{C}\hat{U}_{f}^{l}$ will be applied.
It will implement local unitary operator $U^l_{f(x)}$ on the edge states according to the two extra states.
After doing this, we will reset the extra states $|l\rangle \otimes |f(l)\rangle $ to $|0\rangle \otimes |0\rangle $ for the next run.
This is the explicit implementation of one step of this algorithm and can be realized by quantum circuit \cite{reitzner2009quantum}.

We study the search on the complete graph with $N$ vertices. Each vertex is connected with other vertices, so the dimension of the Hilbert space is $N(N-1)$ and $|\Gamma(l)|=N-1$ for any vertex. The initial state of the walker is the equal superposition of all edge states, i.e.
\begin{equation}
|\psi_{0}\rangle=\frac{1}{\sqrt{N(N-1)}} \sum_{a=1}^{N} \sum_{b=1,a\neq b}^{N} |a,b\rangle .
\end{equation}
Considering the symmetry of the graph, the quantum walk on $N(N-1)$ dimension Hilbert space can be reduced to unitary evolution in a much smaller space consisting of four vectors \cite{reitzner2009quantum}. This is achieved by decomposing $\mathcal{H}$ into four invariant subspaces under all automorphism mapping, i.e. $ \mathcal{H}= \bigoplus_{j=1}^{4} \mathcal{H}_{j}$. Here
\begin{eqnarray}
\mathcal{H}_1&=&l^2(\{|m,l\rangle \ m \in \mathcal{V} ,l \in \mathcal{N},ml \in E  \}), \\ \nonumber
\mathcal{H}_2&=&l^2(\{|m,l\rangle \ m \in \mathcal{N}, l \in \mathcal{V},ml \in E  \}), \\ \nonumber
\mathcal{H}_3&=&l^2(\{|m,l\rangle \ m,l \in \mathcal{N},ml \in E  \}), \\ \nonumber
\mathcal{H}_4&=&l^2(\{|m,l\rangle \ m,l \in \mathcal{V},ml \in E  \}). \\ \nonumber
\end{eqnarray}
The marked vertices are labeled as $1,2,\hdots,v$ and normal vertices are labeled as $v+1,v+2,\hdots, N$. As stated in \cite{reitzner2009quantum}, the vectors constructed from the invariant subspaces are the equal superposition of the states in each subspace, i.e.
\begin{eqnarray}
 |W_1\rangle&=&\frac{1}{\sqrt{v(N-v)}} \sum_{a=v+1}^{N} \sum_{b=1}^{v} |a,b\rangle, \\ \nonumber
 |W_2\rangle&=&\frac{1}{\sqrt{v(N-v)}} \sum_{a=1}^{v} \sum_{b=v+1}^{N} |a,b\rangle,  \\ \nonumber
 |W_3\rangle&=&\frac{1}{\sqrt{(N-v)(N-v-1)}} \sum_{a=v+1}^{N} \sum^{N}_{b=v+1,a\neq b}|a,b\rangle,  \\ \nonumber
 |W_4\rangle&=&\frac{1}{\sqrt{v(v-1)}} \sum_{a=1}^{v} \sum_{b=1,a\neq b}^{v} |a,b\rangle.
\end{eqnarray}
Note that if $v=1$, $\mathcal{H}_4$ will not exist, leading to the absence of $|W_4\rangle$.

Suppose $v>1$, the initial state can be written in the terms of these four vectors
\begin{align}
|\psi_{0}\rangle=&\!\sqrt{\frac{v(N-v)}{N(N-1)}}(|W_1\rangle + |W_2\rangle ) \\ \nonumber
                   &\!+\sqrt{\frac{(N\!-\!v)\!(N\!-\!v\!-\!1)}{N(N-1)}} |W_3\rangle\!+\!\sqrt{\frac{v(v-1)}{N(N-1)}} |W_4\rangle.
\end{align}
The unitary operator of the walker can be represented as
\begin{equation}
U=
\left(
  \begin{array}{cccc}
    0 & q & s & 0 \\
    e^{i \varphi} & 0 & 0 & 0 \\
    0 & s & -q & 0 \\
    0 & 0 & 0 & e^{i \varphi} \\
  \end{array}
\right),
\end{equation}
where
\begin{eqnarray}
q&=&-r+(v-1)t=-1+\frac{2v}{N-1}, \\ \nonumber s&=&\sqrt{1-q^2}=t \sqrt{v(N-v-1)} .
\end{eqnarray}
The \cite{reitzner2009quantum} demonstrated that the efficiency of the algorithm will reach its maximum when the phase shift $\varphi$ is set to $\pi$. In this case, at any time the state of the walker is
\begin{equation}
|\psi_n \rangle
\!=\!D
\left(\!
  \begin{array}{c}
    \sqrt{\!2 (\!N \! - \!1)}\sin(2n+1)\frac{\theta}{2}\!+\!(-1)^{n}A\\
    -\sqrt{\!2 (\!N \!- \!1)}\sin(2n-1)\frac{\theta}{2}\!+\!(-1)^{n}A \\
    2\sqrt{\!(N \!- \! v \! - \!1)}\cos n \theta -(-1)^{n}B\\
    (-1)^{n}C  \\
  \end{array}
  \!
\right),
\end{equation}
where
\begin{eqnarray}
&A=\sqrt{\frac{v(N-v-1)^2}{(N-1)^2}} ,
B=\sqrt{\frac{v^2(N-v-1)}{(N-1)^2}}, \\ \nonumber
&C=\sqrt{\frac{v(v-1)(2N-v-2)^2}{(N-v)(N-1)^2}} ,
D=\sqrt{\frac{(N-v)(N-1)}{(2N-v-2)^{2} N}} ,\\ \nonumber
&\tan\theta=\frac{\sqrt{v(2N-v-2)}}{N-v-1}.
\end{eqnarray}
When $1< v \ll N$, $|\psi_n\rangle $ is reduced to
\begin{equation}
|\psi_n \rangle
=\frac{1}{2}
\left(
  \begin{array}{c}
    \sqrt{2}\sin(2n+1)\frac{\theta}{2} \\
    -\sqrt{2}\sin(2n-1)\frac{\theta}{2} \\
    2 \cos n \theta \\
    0 \\
  \end{array}
\right).
\end{equation}

After measurement, if the walker is standing on the edge connected to a marked vertex, a target is found successfully. The probability of finding a state $|\psi_n\rangle$ on an edge connected to only normal vertices is $|\langle W_3|\psi_n\rangle|^2$, which is also the probability that we fail to find the marked vertices. Thus probability of finding the marked vertices is
\begin{equation}
P_s=1-|\langle W_3|\psi_n\rangle|^2 =\sin^2 n \theta.
\end{equation}
Since $n \theta=\frac{\pi}{2}$, $P_s=1$, the proper time to measure the walker is $[\frac{\pi}{2} \sqrt{\frac{N}{2v}}]$. When $v=1$, $|\psi_n\rangle$ will not have the fourth component, but the evolution of $P_s$ and the proper time to measure the walker will not change.

In \cite{baumgratz2014quantifying}, two functions are proven to be suitable measures of coherence. One is the distance measure based on relative entropy and another one is the $l_1$ norm, which are denoted as $C_r(\hat{\rho})$ and $C_l(\hat{\rho})$, respectively. The explicit expressions of them are
\begin{eqnarray}\label{measure}
C_r(\hat{\rho})&=&S(\hat{\rho}_{diag})-S(\hat{\rho}) \\ \nonumber
&=&-{\rm Tr}(\hat{\rho}_{diag}\log_{2}\hat{\rho}_{diag}-\hat{\rho} \log_{2}\hat{\rho}), \\ \nonumber
C_l(\hat{\rho})&=&\sum_{i,j,i \neq j} |\rho_{ij}| ,
\end{eqnarray}
where $\hat{\rho}$ is the density matrix of the walker, $\hat{\rho}_{diag}$ is the matrix only having the diagonal elements of $\hat{\rho}$ and $\rho_{ij}$ are the entries of the density matrix. At any time, the density matrix of the walker is
\begin{equation}
\hat{\rho}_n=| \psi_n\rangle \langle \psi_n|.
\end{equation}
Note that the state of the walker is a pure state, so the $C_r(\hat{\rho})$ is reduced to $S(\hat{\rho}_{diag})$.
Applying Eq.(\ref{measure}) to the state of the walker and considering  $1<v \ll N$, we have that
\begin{align}\label{Cr}
C_r(\hat{\rho}_n)=& H (\sin^2n\theta)\!+\!\cos^2n \theta\log_2N^2\!+\!\sin^2n \theta\log_2 2 N\!v  \nonumber \\
                &-\frac{v(v-1)}{N(N-1)}\log_2 \frac{1}{N(N-1)}
\end{align}
and
\begin{equation}\label{Cl}
C_l(\hat{\rho}_n)\!=\!2\!N\!v\!\sin^2\!(n\theta)\!+\!N^2\!\!\cos^2\!n \theta \!+\!\sqrt{2\!N\!v}N|\!\sin(2n\theta)\!|,
\end{equation}
where $H(x)$ is the binary Shannon entropy.

From the above discussions, we can see that coherence and probability of success are all periodic. To reach a high efficiency (evaluated by queries of oracles), it is reasonable for us to measure the state before or when probability of success reaches its maximum. So our discussion about probability of success and coherence is confined to the half-period until probability of success reaches its maximum. The first term in $C_r(\hat{\rho})$ is the binary Shannon entropy which is smaller than $1$ and the sum of second term and third term is monotonically decrease before the probability of success reaches its maximum. Note that $N \gg v >1$, the dynamic of coherence is governed by the sum of second term and third term in Eq. (\ref{Cr}) under $C_r(\hat{\rho})$ and second term in Eq. (\ref{Cl}) under $C_l(\hat{\rho})$, so the coherence of the walker will decrease monotonically before the probability of success reaches its maximum.
\begin{figure}[htbp]
 {
 \includegraphics[width=0.5\textwidth]{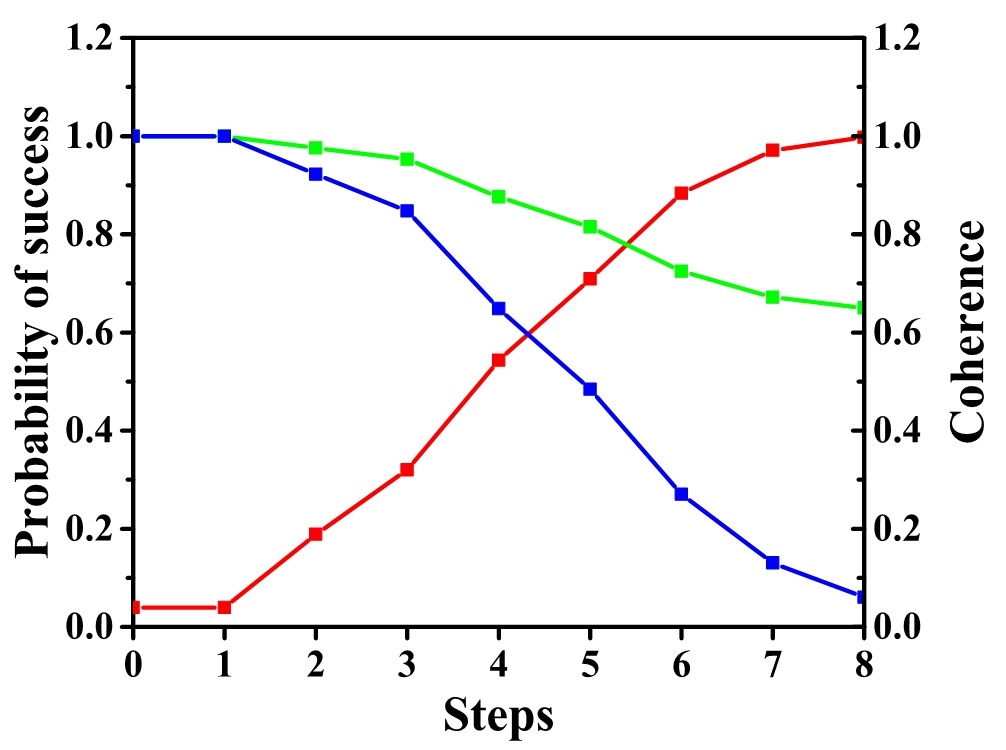}
 }
 \caption{The number of total vertices is 100 and the number of marked vertices is 2. The phase shift is set to $\pi$. The red line is the probability of success. The green line is the coherence under $C_r(\hat{\rho})$ and the blue line is the coherence under $C_l(\hat{\rho})$. The values of two measures of coherence are normalized to 1. The X axis is the step of implementation of this algorithm. The left Y axis is the probability of success and right Y axis is the value of normalized measure of coherence.}\label{Graph1and2}
\end{figure}
It shows that the walker has consumed the coherence of the initial state to complete the task of search. We define the depletion of coherence from the initial state to the state with maximal probability of success as the consumption of coherence. The connection between the coherence and the probability of success implies that coherence should be viewed as a resource in this algorithm. The results of analysis are also supported by numerical calculation presented in Fig. \ref{Graph1and2}.

For Grover search algorithm, similar result has been obtained recently \cite{shi2016coherence}. Since the quantum algorithm we studied is defined by local unitary operator, it is different from Grover search.
For one particle quantum walk search, the state of walker is a single quantum state with multi levels,
the methods of quantum entanglement and quantum correlations are in general not applicable,
it is expected that the coherence is the resource in this algorithm.
If we choose single quantum state with multi-levels rather than multi qubits to build quantum database and replace Walsh-Hadamard $H^{\otimes n}$ with $U$ which transforms $|0\rangle $ to $\frac{1}{\sqrt{N}} \sum_{i=0}^{N-1} |i\rangle$, the situation will be the same in Grover search.
\section{Connection between the efficiency of the algorithm and coherence} 
\label{sec:reduce_efficiency}
\begin{figure}
 {
 \includegraphics[width=0.5\textwidth]{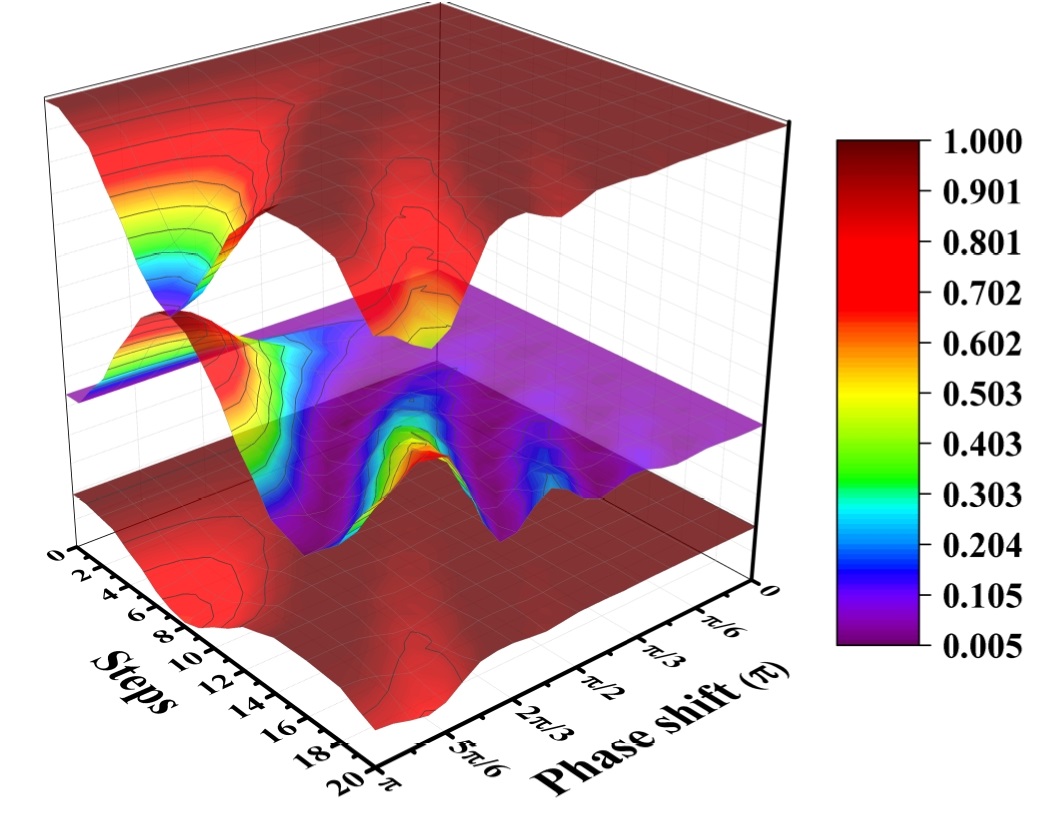}
 }
 \caption{The total number of the vertices is set to 100 and the number of marked vertices is set to 2. The middle surface is the probability of success. The bottom surface and top surface are the coherence under $C_r(\hat{\rho})$ and $C_l(\hat{\rho})$, respectively. The values of two measures of coherence are normalized to 1.}\label{Graph3and4}
 \end{figure}

\begin{figure}
 {
 \includegraphics[width=0.5\textwidth]{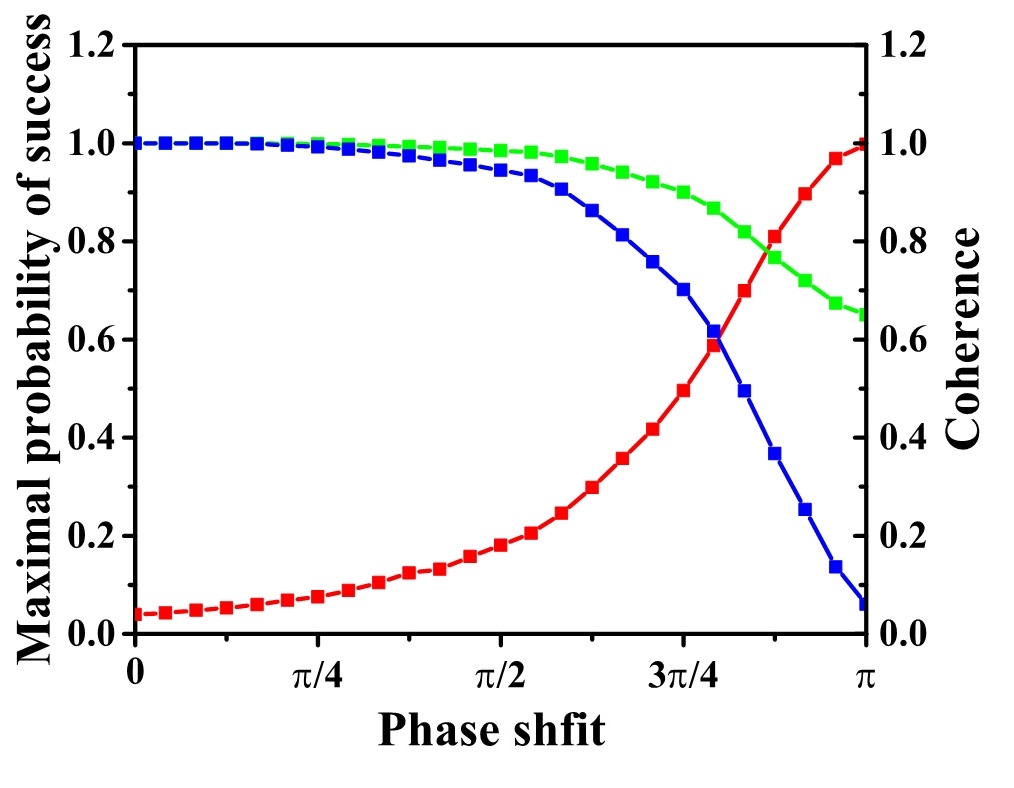}
 }
 \caption{$N=100$, $v=2$. The red line is the maximal probability of success. The green line is the coherence under $C_r(\hat{\rho})$ and the blue line is the coherence under $C_l(\hat{\rho})$. The values of two measures of coherence are normalized to 1. The X axis is the angle of phase shift. Here we choose 25 values evenly in the interval from $0$ to $\pi$. The left Y axis is the maximal probability of success and right Y axis is the value of normalized measure of coherence.}\label{Graph5and6}
 \end{figure}

The \cite{reitzner2009quantum} shows that when $\varphi$ is not $\pi$, the maximal probability will not be unit,  indicating the decrease of the efficiency of the algorithm. We numerically calculated the dynamics of coherence of the processes with different $\varphi$. When $\varphi$ is substituted by $\pi- \varphi$, $U$ will change into $U^*$. Note that $|\psi_{0}\rangle$ is real, it is easy to see that $(U^*)^{n} |\psi_{0}\rangle=(U^n|\psi_{0}\rangle)^*$. Then the norm of the amplitude of the state at any time will be the same when $\varphi$ is changed to $\pi- \varphi$. Further more, the probability of success and coherence will not change. So we only consider $\varphi$ in the region $[0,\pi]$. The value of coherence and probability of success are presented in Fig.\ref{Graph3and4}.
As we can see from the figure, when $\varphi$ is no longer $\pi$, the correspondence between the success probability and the coherence is still preserved.
However, the consumption of coherence decreases with lower maximal probability of success.
With very low probability of success, the coherence approaches the unit.
When phase shift is closed to $0$, the probability of success is very low, leading to the false of finding targets. In this process, the coherence is stable.
For a more clear presentation, we give the mean of local maximal probability of success and corresponding minimal coherence at that step for different $\varphi$ in Fig.\ref{Graph5and6}.
In this figure, when maximal probability of success decreases, the minimal coherence increases. This result indicates that when the algorithm is less efficient, the consumption of coherence will decrease.

\begin{figure}
  \centering \includegraphics[width=0.5\textwidth]{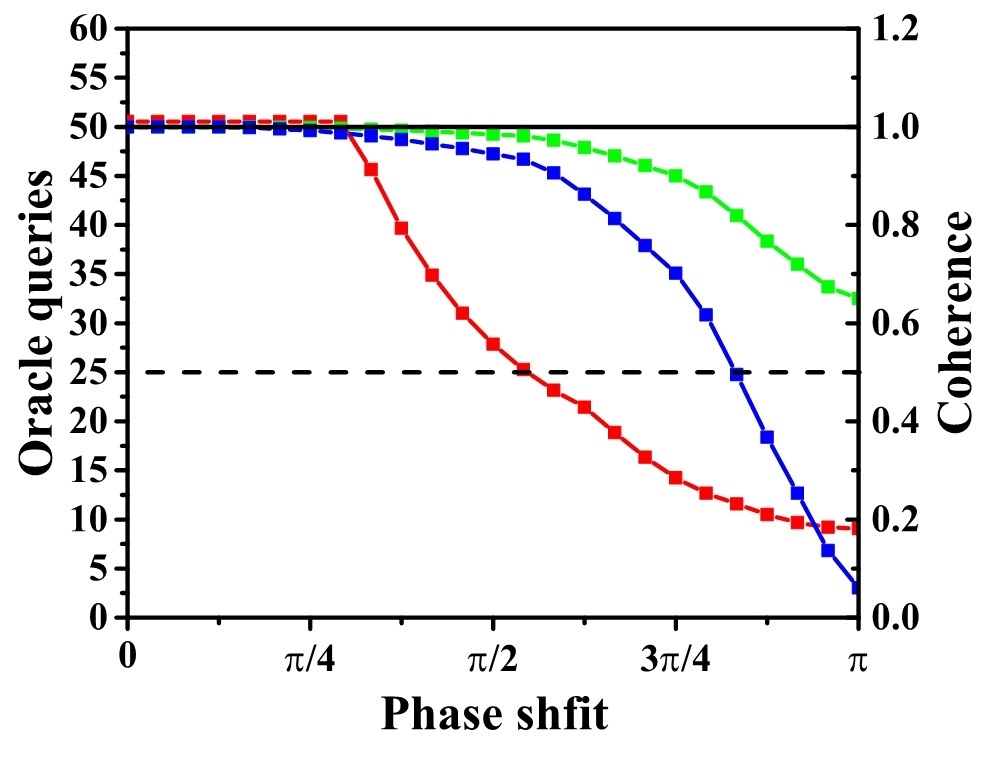}
  \caption{$N=100$, $v=2$. The red line is the number of oracle queries. The green line is the coherence under $C_r(\hat{\rho})$ and the blue line is the coherence under $C_l(\hat\rho)$. The values of two measures of coherence are normalized to 1. The X axis is the angle of phase shift from $0$ to $\pi$. The left Y axis is the average number of oracle queries and right Y axis is the value of normalized measure of coherence. The dash black line is the number of oracle queries of classical search with memory and black solid line is the number of oracle queries of classical blind search.}\label{Graph8}
 \end{figure}

 To show the connection between the consumption of coherence and the efficiency of the algorithm more explicitly, the number of oracle queries is chosen to evaluate its efficiency. The oracle queries can be calculated as follows. Firstly, we will let the particle walk $m$ times, each time it will query for an oracle. Then we will measure the particle and reduce the state to one of the edge states. Two oracles will be used to evaluate whether the two vertices connected to the edge are the marked vertices. Suppose we find the marked vertex at the $k$-th run of the algorithm, the number of oracle queries will be $k(m+2)$. Since the maximal probability of success is not unit, the marked vertices can not always be found by a single run of the algorithm. Then an average number of oracle queries is defined as
 \begin{equation}
 \bar{Q}_{\varphi,m}\!=\!\sum_{k=1}^{\infty} (1\!-\!\!P_{\varphi}(m))^{k\!-\!1}\!P_{\varphi}(m) k(m+1)\!=\!\frac{m+2}{P_{\varphi}(m)},
 \end{equation}
where $P_{\varphi}(m)$ is the probability of success when we measure the walker at $m$-th step. Here, $m$ is chosen to minimize $\bar{Q}_{\varphi,m} $.
This quantity represents the efficiency of the algorithm. With fewer oracle queries, the algorithm will be more efficient.
The result for the case of $100$ vertices with $2$ marked vertices is plotted in Fig. \ref{Graph8}. It shows that the classical blind search provides a lower bound for the efficiency of the quantum search algorithm. The algorithm will be less efficient than the classical search with memory, if the consumption of coherence is below a proper value.
These results tell us that coherence is responsible for the speed up of this quantum search algorithm.

\section{Walk with no coherence} 
\label{sec:walk_with_no_coherence}
In this section, we focus on the probability of success when the initial state is incoherent. According to the \cite{baumgratz2014quantifying}, an incoherent initial state should be expressed as
\begin{equation}
\rho_0= \sum^{N}_{a=1} \sum^{N}_{b=1,a \neq b} P_{ab}|a,b\rangle \langle a,b|,
\end{equation}
where $\sum_{a,b}P_{ab}=1$.
At any time, density matrix of the system is
\begin{equation}
\rho(n)=U^n \rho_0 U^{\dag n}.
\end{equation}
This can also be viewed as that $U$ is performed $n$ times on ensemble $\{|a,b\rangle \}$ with probability $P_{ab}$ for the state $|a,b\rangle$.
With different states in the ensemble as initial state, the probability of success for state $|a,b\rangle$ is
\begin{equation}\label{success}
 P_s(|a,b\rangle,n)=1- \sum_{k=v+1}^{N} \sum_{l=v+1,l \neq v}^{N} |\langle k,l | U^n|a,b\rangle |^2.
\end{equation}
The success probability for an initial incoherent state would be
\begin{equation}
P_s(n)=\sum^{N}_{a=1} \sum^{N}_{b=1,a \neq b} P_s(|a,b\rangle ,n)P_{ab}.
\end{equation}
The Eq.(\ref{success}) shows that, if the state comes from the same subspace, they will share the same value of $P_s(|a,b\rangle,n)$. Then we define $P_s(\mathcal{H}_i,n)$, which is the probability of success with an initial state $|a,b \rangle \in \mathcal{H}_i$. As we stated above, there are four different subspaces (number of marked ones is more than 1). In the case of an incoherent initial state, the probability of success can be reformulated as
\begin{eqnarray}
P_s(n)=\sum_{i=1}^{4} a_iP_s(\mathcal{H}_i,n),
\end{eqnarray}
where
\begin{eqnarray}
a_i&=&\sum_{|a,b\rangle \in \mathcal{H}_i} P_{ab}.
\end{eqnarray}

\begin{figure}
 \centering \includegraphics[width=0.5\textwidth]{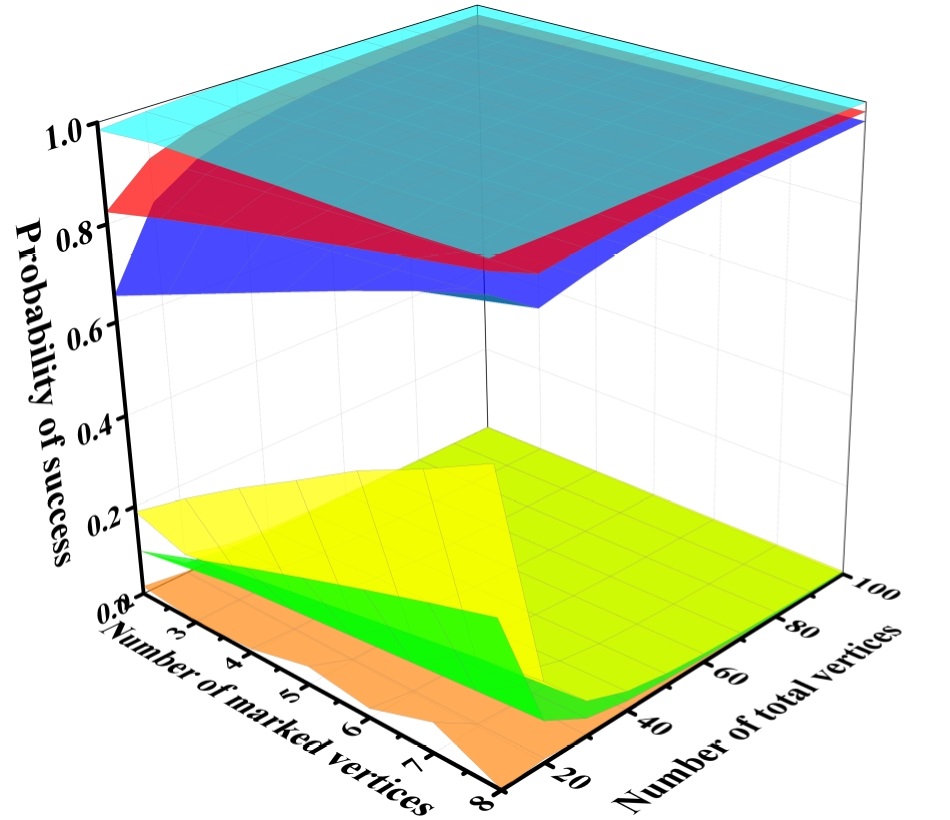}
 \caption{The red surface and green surface are the means of $P_s(\mathcal{H}_1,n)$ and $P_s(\mathcal{H}_3,n)$, respectively. The surfaces above and below the red and green surface are the maximal and minimal values of $P_s(\mathcal{H}_1,n)$ and $P_s(\mathcal{H}_3,n)$ respectively with different $N$ and $v$. The Z axis is the probability of success.}\label{Graph4}
\end{figure}

Note that at the beginning, $P_s(\mathcal{H}_3,0)=0$, it is easy to obtain that $a_1+a_2+a_4=P_s(0)$. For a state from the subspace $\mathcal{H}_4$, it will transform between $|a,b\rangle$ and $|b,a\rangle$ with additional minus with implementation of $U$. Thus $P_s(\mathcal{H}_4,n)$ will be 1 all the time. When we apply $U$ on the state from $\mathcal{H}_2$, it will turn into one from $\mathcal{H}_1$ with additional phase shift $\pi$. Thus for the states from subspace $\mathcal{H}_1$ and $\mathcal{H}_2 $, it is easy to obtain that $P_s(\mathcal{H}_1,n)=P_s(\mathcal{H}_2,n+1)$ with $P_s(\mathcal{H}_2,0)=1$. We numerically calculate $P_s(n)$ for states from subspace $\mathcal{H}_1$ and $\mathcal{H}_3$ and present the result in Fig .\ref{Graph4}.
Then we can conclude that when $N \gg v$, $P_s(\mathcal{H}_1,n)$, $P_s(\mathcal{H}_2,n)$ will converge to one and $P_s(\mathcal{H}_3, n)$ will be zero. This result is very reasonable. In another scheme of quantum search, quantum amplitude amplification, if $A|0 \rangle $ is totally projected on the good subspace or bad subspace, the probability of success will be 1 or 0, respectively \cite{brassard2002quantum}. However, this only holds for $N \gg v$ in quantum scattering walk search. Recall that $P_s(\mathcal{H}_4,n)$ is $1$ all the time, the probability of success will be
\begin{equation}
P_s(n)=a_1+a_2+a_4=P_s(0).
\end{equation}
The result above implies that the probability of success is only determined by the initial state $\rho_0$. It means that the scattering quantum walk search totally loses its power on all incoherent initial states and further shows that coherence should be considered as a resource in this algorithm.
\section{Conclusion} 
\label{sec:conclusion}

In this paper, we calculate the coherence in the scattering quantum walk search algorithm on complete graph and connect it with the probability of success under the condition $N \gg v >1$.
We found that the coherence of the system decreases while the probability of success increases until reaching its maximum.
It shows that coherence of the system is consumed to complete the task of search.
Since the system we considered is a single quantum state with multi levels,
the methods of quantum entanglement and quantum correlations are in general not applicable,
the coherence clearly can be considered as the resource in this algorithm.
Besides, with the value of phase shift varied, the decrease of maximal success probability causes the simultaneous decline of coherence consumption, which gives rise to the potential correspondence between the efficiency of the algorithm and the consumption of coherence. Choosing oracle queries to evaluate the efficiency of the algorithm, we find the consumption of coherence varies as an monotonically increasing function of the efficiency of algorithm. If the coherence consumption is smaller than a given value, the quantum algorithm we investigated will be less efficient than the classical search with memory. Without coherence consumption, the quantum walk search algorithm will have the same efficiency as the classical blind search. That is to say, the coherence is responsible for the speed up of this quantum algorithm. Last but not the least, we consider the probability of success of this algorithm starting with an incoherent state and discover that it keeps unchanged compared with the initial time. It clearly shows that the coherence is a key resource in this algorithm.

Our work shows that coherence plays an essential role and is responsible for the speed up in scattering quantum walk search on complete graph. We believe that our method can be generalized to other quantum walk algorithms and may have applications in the quantum computation processing.

\begin{acknowledgments}
We thank D. Reitzner and W. Dong for their valuable discussions. This work was supported by the NSFC (Grant No.11375141, No.11425522, No.91536108 and No.11647057), the special research funds of shaanxi province department of education (No.203010005), Northwest University scientific research funds (No.338020004) and the double first-class university construction project of Northwest University.
\end{acknowledgments}
\bibliography{ref}

\begin{thebibliography}{33}%
\makeatletter
\providecommand \@ifxundefined [1]{%
 \@ifx{#1\undefined}
}%
\providecommand \@ifnum [1]{%
 \ifnum #1\expandafter \@firstoftwo
 \else \expandafter \@secondoftwo
 \fi
}%
\providecommand \@ifx [1]{%
 \ifx #1\expandafter \@firstoftwo
 \else \expandafter \@secondoftwo
 \fi
}%
\providecommand \natexlab [1]{#1}%
\providecommand \enquote  [1]{``#1''}%
\providecommand \bibnamefont  [1]{#1}%
\providecommand \bibfnamefont [1]{#1}%
\providecommand \citenamefont [1]{#1}%
\providecommand \href@noop [0]{\@secondoftwo}%
\providecommand \href [0]{\begingroup \@sanitize@url \@href}%
\providecommand \@href[1]{\@@startlink{#1}\@@href}%
\providecommand \@@href[1]{\endgroup#1\@@endlink}%
\providecommand \@sanitize@url [0]{\catcode `\\12\catcode `\$12\catcode
  `\&12\catcode `\#12\catcode `\^12\catcode `\_12\catcode `\%12\relax}%
\providecommand \@@startlink[1]{}%
\providecommand \@@endlink[0]{}%
\providecommand \url  [0]{\begingroup\@sanitize@url \@url }%
\providecommand \@url [1]{\endgroup\@href {#1}{\urlprefix }}%
\providecommand \urlprefix  [0]{URL }%
\providecommand \Eprint [0]{\href }%
\providecommand \doibase [0]{http://dx.doi.org/}%
\providecommand \selectlanguage [0]{\@gobble}%
\providecommand \bibinfo  [0]{\@secondoftwo}%
\providecommand \bibfield  [0]{\@secondoftwo}%
\providecommand \translation [1]{[#1]}%
\providecommand \BibitemOpen [0]{}%
\providecommand \bibitemStop [0]{}%
\providecommand \bibitemNoStop [0]{.\EOS\space}%
\providecommand \EOS [0]{\spacefactor3000\relax}%
\providecommand \BibitemShut  [1]{\csname bibitem#1\endcsname}%
\let\auto@bib@innerbib\@empty
\bibitem [{\citenamefont {Motwani}\ and\ \citenamefont
  {Raghavan}(1995)}]{motwani1995}%
  \BibitemOpen
  \bibfield  {author} {\bibinfo {author} {\bibfnamefont {R.}~\bibnamefont
  {Motwani}}\ and\ \bibinfo {author} {\bibfnamefont {P.}~\bibnamefont
  {Raghavan}},\ }\href@noop {} {\emph {\bibinfo {title} {Randomized
  Algorithms}}}\ (\bibinfo  {publisher} {Cambridge University Press},\ \bibinfo
  {address} {Cambridge, UK},\ \bibinfo {year} {1995})\BibitemShut {NoStop}%
\bibitem [{\citenamefont {Schoning}(1999)}]{schoning1999probabilistic}%
  \BibitemOpen
  \bibfield  {author} {\bibinfo {author} {\bibfnamefont {T.}~\bibnamefont
  {Schoning}},\ }in\ \href@noop {} {\emph {\bibinfo {booktitle} {Proceedings of
  the 40th Annual Symposium on Foundations of Computer Science}}}\ (\bibinfo
  {organization} {IEEE},\ \bibinfo {address} {New York},\ \bibinfo {year}
  {1999})\ p.\ \bibinfo {pages} {410}\BibitemShut {NoStop}%
\bibitem [{\citenamefont {Aharonov}\ \emph {et~al.}(1993)\citenamefont
  {Aharonov}, \citenamefont {Davidovich},\ and\ \citenamefont
  {Zagury}}]{Quantum_random_walks}%
  \BibitemOpen
  \bibfield  {author} {\bibinfo {author} {\bibfnamefont {Y.}~\bibnamefont
  {Aharonov}}, \bibinfo {author} {\bibfnamefont {L.}~\bibnamefont
  {Davidovich}}, \ and\ \bibinfo {author} {\bibfnamefont {N.}~\bibnamefont
  {Zagury}},\ }\href {\doibase 10.1103/PhysRevA.48.1687} {\bibfield  {journal}
  {\bibinfo  {journal} {Phys. Rev. A}\ }\textbf {\bibinfo {volume} {48}},\
  \bibinfo {pages} {1687} (\bibinfo {year} {1993})}\BibitemShut {NoStop}%
\bibitem [{\citenamefont {Shor}(1997)}]{shor1999polynomial}%
  \BibitemOpen
  \bibfield  {author} {\bibinfo {author} {\bibfnamefont {P.~W.}\ \bibnamefont
  {Shor}},\ }\href {\doibase 10.1137/S0036144598347011} {\bibfield  {journal}
  {\bibinfo  {journal} {SIAM J. Comput.}\ }\textbf {\bibinfo {volume} {26}},\
  \bibinfo {pages} {1484} (\bibinfo {year} {1997})}\BibitemShut {NoStop}%
\bibitem [{\citenamefont {Grover}(1997)}]{grover1997quantum}%
  \BibitemOpen
  \bibfield  {author} {\bibinfo {author} {\bibfnamefont {L.~K.}\ \bibnamefont
  {Grover}},\ }\href {\doibase 10.1103/PhysRevLett.79.325} {\bibfield
  {journal} {\bibinfo  {journal} {Phys. Rev. Lett.}\ }\textbf {\bibinfo
  {volume} {79}},\ \bibinfo {pages} {325} (\bibinfo {year} {1997})}\BibitemShut
  {NoStop}%
\bibitem [{\citenamefont {Ambainis}\ \emph {et~al.}(2001)\citenamefont
  {Ambainis}, \citenamefont {Bach}, \citenamefont {Nayak}, \citenamefont
  {Vishwanath},\ and\ \citenamefont {Watrous}}]{ambainis2001one}%
  \BibitemOpen
  \bibfield  {author} {\bibinfo {author} {\bibfnamefont {A.}~\bibnamefont
  {Ambainis}}, \bibinfo {author} {\bibfnamefont {E.}~\bibnamefont {Bach}},
  \bibinfo {author} {\bibfnamefont {A.}~\bibnamefont {Nayak}}, \bibinfo
  {author} {\bibfnamefont {A.}~\bibnamefont {Vishwanath}}, \ and\ \bibinfo
  {author} {\bibfnamefont {J.}~\bibnamefont {Watrous}},\ }in\ \href {\doibase
  10.1145/380752.380757} {\emph {\bibinfo {booktitle} {Proceedings of the 33rd
  ACM STOC}}}\ (\bibinfo  {publisher} {ACM},\ \bibinfo {address} {New York},\
  \bibinfo {year} {2001})\ p.~\bibinfo {pages} {37}\BibitemShut {NoStop}%
\bibitem [{\citenamefont {Childs}\ \emph {et~al.}(2003)\citenamefont {Childs},
  \citenamefont {Cleve}, \citenamefont {Deotto}, \citenamefont {Farhi},
  \citenamefont {Gutmann},\ and\ \citenamefont
  {Spielman}}]{childs2003exponential}%
  \BibitemOpen
  \bibfield  {author} {\bibinfo {author} {\bibfnamefont {A.~M.}\ \bibnamefont
  {Childs}}, \bibinfo {author} {\bibfnamefont {R.}~\bibnamefont {Cleve}},
  \bibinfo {author} {\bibfnamefont {E.}~\bibnamefont {Deotto}}, \bibinfo
  {author} {\bibfnamefont {E.}~\bibnamefont {Farhi}}, \bibinfo {author}
  {\bibfnamefont {S.}~\bibnamefont {Gutmann}}, \ and\ \bibinfo {author}
  {\bibfnamefont {D.~A.}\ \bibnamefont {Spielman}},\ }in\ \href {\doibase
  10.1145/780542.780552} {\emph {\bibinfo {booktitle} {Proceedings of the 35th
  ACM STOC}}}\ (\bibinfo  {publisher} {ACM},\ \bibinfo {address} {New York},\
  \bibinfo {year} {2003})\ p.~\bibinfo {pages} {59}\BibitemShut {NoStop}%
\bibitem [{\citenamefont {Shenvi}\ \emph {et~al.}(2003)\citenamefont {Shenvi},
  \citenamefont {Kempe},\ and\ \citenamefont {Whaley}}]{shenvi2003quantum}%
  \BibitemOpen
  \bibfield  {author} {\bibinfo {author} {\bibfnamefont {N.}~\bibnamefont
  {Shenvi}}, \bibinfo {author} {\bibfnamefont {J.}~\bibnamefont {Kempe}}, \
  and\ \bibinfo {author} {\bibfnamefont {K.~B.}\ \bibnamefont {Whaley}},\
  }\href {\doibase 10.1103/PhysRevA.67.052307} {\bibfield  {journal} {\bibinfo
  {journal} {Phys. Rev. A}\ }\textbf {\bibinfo {volume} {67}},\ \bibinfo
  {pages} {052307} (\bibinfo {year} {2003})}\BibitemShut {NoStop}%
\bibitem [{\citenamefont {Childs}\ and\ \citenamefont
  {Goldstone}(2004)}]{childs2004spatial}%
  \BibitemOpen
  \bibfield  {author} {\bibinfo {author} {\bibfnamefont {A.~M.}\ \bibnamefont
  {Childs}}\ and\ \bibinfo {author} {\bibfnamefont {J.}~\bibnamefont
  {Goldstone}},\ }\href {\doibase 10.1103/PhysRevA.70.042312} {\bibfield
  {journal} {\bibinfo  {journal} {Phys. Rev. A}\ }\textbf {\bibinfo {volume}
  {70}},\ \bibinfo {pages} {042312} (\bibinfo {year} {2004})}\BibitemShut
  {NoStop}%
\bibitem [{\citenamefont {Schmitz}\ \emph {et~al.}(2009)\citenamefont
  {Schmitz}, \citenamefont {Matjeschk}, \citenamefont {Schneider},
  \citenamefont {Glueckert}, \citenamefont {Enderlein}, \citenamefont {Huber},\
  and\ \citenamefont {Schaetz}}]{schmitz2009quantum}%
  \BibitemOpen
  \bibfield  {author} {\bibinfo {author} {\bibfnamefont {H.}~\bibnamefont
  {Schmitz}}, \bibinfo {author} {\bibfnamefont {R.}~\bibnamefont {Matjeschk}},
  \bibinfo {author} {\bibfnamefont {C.}~\bibnamefont {Schneider}}, \bibinfo
  {author} {\bibfnamefont {J.}~\bibnamefont {Glueckert}}, \bibinfo {author}
  {\bibfnamefont {M.}~\bibnamefont {Enderlein}}, \bibinfo {author}
  {\bibfnamefont {T.}~\bibnamefont {Huber}}, \ and\ \bibinfo {author}
  {\bibfnamefont {T.}~\bibnamefont {Schaetz}},\ }\href {\doibase
  10.1103/PhysRevLett.103.090504} {\bibfield  {journal} {\bibinfo  {journal}
  {Phys. Rev. Lett.}\ }\textbf {\bibinfo {volume} {103}},\ \bibinfo {pages}
  {090504} (\bibinfo {year} {2009})}\BibitemShut {NoStop}%
\bibitem [{\citenamefont {Xue}\ \emph {et~al.}(2009)\citenamefont {Xue},
  \citenamefont {Sanders},\ and\ \citenamefont {Leibfried}}]{xue2009quantum}%
  \BibitemOpen
  \bibfield  {author} {\bibinfo {author} {\bibfnamefont {P.}~\bibnamefont
  {Xue}}, \bibinfo {author} {\bibfnamefont {B.~C.}\ \bibnamefont {Sanders}}, \
  and\ \bibinfo {author} {\bibfnamefont {D.}~\bibnamefont {Leibfried}},\ }\href
  {\doibase 10.1103/PhysRevLett.103.183602} {\bibfield  {journal} {\bibinfo
  {journal} {Phys. Rev. Lett.}\ }\textbf {\bibinfo {volume} {103}},\ \bibinfo
  {pages} {183602} (\bibinfo {year} {2009})}\BibitemShut {NoStop}%
\bibitem [{\citenamefont {Peruzzo}\ \emph {et~al.}(2010)\citenamefont
  {Peruzzo}, \citenamefont {Lobino}, \citenamefont {Matthews} \emph
  {et~al.}}]{peruzzo2010quantum}%
  \BibitemOpen
  \bibfield  {author} {\bibinfo {author} {\bibfnamefont {A.}~\bibnamefont
  {Peruzzo}}, \bibinfo {author} {\bibfnamefont {M.}~\bibnamefont {Lobino}},
  \bibinfo {author} {\bibfnamefont {J.~C.}\ \bibnamefont {Matthews}},  \emph
  {et~al.},\ }\href {\doibase 10.1126/science.1193515} {\bibfield  {journal}
  {\bibinfo  {journal} {Science}\ }\textbf {\bibinfo {volume} {329}},\ \bibinfo
  {pages} {1500} (\bibinfo {year} {2010})}\BibitemShut {NoStop}%
\bibitem [{\citenamefont {Z{\"a}hringer}\ \emph {et~al.}(2010)\citenamefont
  {Z{\"a}hringer}, \citenamefont {Kirchmair}, \citenamefont {Gerritsma},
  \citenamefont {Solano}, \citenamefont {Blatt},\ and\ \citenamefont
  {Roos}}]{zahringer2010realization}%
  \BibitemOpen
  \bibfield  {author} {\bibinfo {author} {\bibfnamefont {F.}~\bibnamefont
  {Z{\"a}hringer}}, \bibinfo {author} {\bibfnamefont {G.}~\bibnamefont
  {Kirchmair}}, \bibinfo {author} {\bibfnamefont {R.}~\bibnamefont
  {Gerritsma}}, \bibinfo {author} {\bibfnamefont {E.}~\bibnamefont {Solano}},
  \bibinfo {author} {\bibfnamefont {R.}~\bibnamefont {Blatt}}, \ and\ \bibinfo
  {author} {\bibfnamefont {C.}~\bibnamefont {Roos}},\ }\href {\doibase
  10.1103/PhysRevLett.104.100503} {\bibfield  {journal} {\bibinfo  {journal}
  {Phys. Rev. Lett.}\ }\textbf {\bibinfo {volume} {104}},\ \bibinfo {pages}
  {100503} (\bibinfo {year} {2010})}\BibitemShut {NoStop}%
\bibitem [{\citenamefont {Xue}\ \emph {et~al.}(2014)\citenamefont {Xue},
  \citenamefont {Qin},\ and\ \citenamefont {Tang}}]{xue2014trapping}%
  \BibitemOpen
  \bibfield  {author} {\bibinfo {author} {\bibfnamefont {P.}~\bibnamefont
  {Xue}}, \bibinfo {author} {\bibfnamefont {H.}~\bibnamefont {Qin}}, \ and\
  \bibinfo {author} {\bibfnamefont {B.}~\bibnamefont {Tang}},\ }\href {\doibase
  10.1038/srep04825} {\bibfield  {journal} {\bibinfo  {journal} {Sci. Rep.}\
  }\textbf {\bibinfo {volume} {4}},\ \bibinfo {pages} {4825} (\bibinfo {year}
  {2014})}\BibitemShut {NoStop}%
\bibitem [{\citenamefont {Bian}\ \emph {et~al.}(2015)\citenamefont {Bian},
  \citenamefont {Li}, \citenamefont {Qin}, \citenamefont {Zhan}, \citenamefont
  {Zhang}, \citenamefont {Sanders},\ and\ \citenamefont
  {Xue}}]{bian2015realization}%
  \BibitemOpen
  \bibfield  {author} {\bibinfo {author} {\bibfnamefont {Z.}~\bibnamefont
  {Bian}}, \bibinfo {author} {\bibfnamefont {J.}~\bibnamefont {Li}}, \bibinfo
  {author} {\bibfnamefont {H.}~\bibnamefont {Qin}}, \bibinfo {author}
  {\bibfnamefont {X.}~\bibnamefont {Zhan}}, \bibinfo {author} {\bibfnamefont
  {R.}~\bibnamefont {Zhang}}, \bibinfo {author} {\bibfnamefont {B.~C.}\
  \bibnamefont {Sanders}}, \ and\ \bibinfo {author} {\bibfnamefont
  {P.}~\bibnamefont {Xue}},\ }\href {\doibase 10.1103/PhysRevLett.114.203602}
  {\bibfield  {journal} {\bibinfo  {journal} {Phys. Rev. Lett.}\ }\textbf
  {\bibinfo {volume} {114}},\ \bibinfo {pages} {203602} (\bibinfo {year}
  {2015})}\BibitemShut {NoStop}%
\bibitem [{\citenamefont {Xue}\ \emph {et~al.}(2015)\citenamefont {Xue},
  \citenamefont {Zhang}, \citenamefont {Qin}, \citenamefont {Zhan},
  \citenamefont {Bian}, \citenamefont {Li},\ and\ \citenamefont
  {Sanders}}]{xue2015experimental}%
  \BibitemOpen
  \bibfield  {author} {\bibinfo {author} {\bibfnamefont {P.}~\bibnamefont
  {Xue}}, \bibinfo {author} {\bibfnamefont {R.}~\bibnamefont {Zhang}}, \bibinfo
  {author} {\bibfnamefont {H.}~\bibnamefont {Qin}}, \bibinfo {author}
  {\bibfnamefont {X.}~\bibnamefont {Zhan}}, \bibinfo {author} {\bibfnamefont
  {Z.}~\bibnamefont {Bian}}, \bibinfo {author} {\bibfnamefont {J.}~\bibnamefont
  {Li}}, \ and\ \bibinfo {author} {\bibfnamefont {B.~C.}\ \bibnamefont
  {Sanders}},\ }\href {\doibase 10.1103/PhysRevLett.114.140502} {\bibfield
  {journal} {\bibinfo  {journal} {Phys. Rev. Lett.}\ }\textbf {\bibinfo
  {volume} {114}},\ \bibinfo {pages} {140502} (\bibinfo {year}
  {2015})}\BibitemShut {NoStop}%
\bibitem [{\citenamefont {Hillery}\ \emph {et~al.}(2003)\citenamefont
  {Hillery}, \citenamefont {Bergou},\ and\ \citenamefont
  {Feldman}}]{Scattering_quantum_walk}%
  \BibitemOpen
  \bibfield  {author} {\bibinfo {author} {\bibfnamefont {M.}~\bibnamefont
  {Hillery}}, \bibinfo {author} {\bibfnamefont {J.}~\bibnamefont {Bergou}}, \
  and\ \bibinfo {author} {\bibfnamefont {E.}~\bibnamefont {Feldman}},\ }\href
  {\doibase 10.1103/PhysRevA.68.032314} {\bibfield  {journal} {\bibinfo
  {journal} {Phys. Rev. A}\ }\textbf {\bibinfo {volume} {68}},\ \bibinfo
  {pages} {032314} (\bibinfo {year} {2003})}\BibitemShut {NoStop}%
\bibitem [{\citenamefont {Reitzner}\ \emph {et~al.}(2009)\citenamefont
  {Reitzner}, \citenamefont {Hillery}, \citenamefont {Feldman},\ and\
  \citenamefont {Bu{\v{z}}ek}}]{reitzner2009quantum}%
  \BibitemOpen
  \bibfield  {author} {\bibinfo {author} {\bibfnamefont {D.}~\bibnamefont
  {Reitzner}}, \bibinfo {author} {\bibfnamefont {M.}~\bibnamefont {Hillery}},
  \bibinfo {author} {\bibfnamefont {E.}~\bibnamefont {Feldman}}, \ and\
  \bibinfo {author} {\bibfnamefont {V.}~\bibnamefont {Bu{\v{z}}ek}},\ }\href
  {\doibase 10.1103/PhysRevA.79.012323} {\bibfield  {journal} {\bibinfo
  {journal} {Phys. Rev. A}\ }\textbf {\bibinfo {volume} {79}},\ \bibinfo
  {pages} {012323} (\bibinfo {year} {2009})}\BibitemShut {NoStop}%
\bibitem [{\citenamefont {Horodecki}\ \emph {et~al.}(2009)\citenamefont
  {Horodecki}, \citenamefont {Horodecki}, \citenamefont {Horodecki},\ and\
  \citenamefont {Horodecki}}]{horodecki2009quantum}%
  \BibitemOpen
  \bibfield  {author} {\bibinfo {author} {\bibfnamefont {R.}~\bibnamefont
  {Horodecki}}, \bibinfo {author} {\bibfnamefont {P.}~\bibnamefont
  {Horodecki}}, \bibinfo {author} {\bibfnamefont {M.}~\bibnamefont
  {Horodecki}}, \ and\ \bibinfo {author} {\bibfnamefont {K.}~\bibnamefont
  {Horodecki}},\ }\href {\doibase 10.1103/RevModPhys.81.865} {\bibfield
  {journal} {\bibinfo  {journal} {Rev. Mod. Phys.}\ }\textbf {\bibinfo {volume}
  {81}},\ \bibinfo {pages} {865} (\bibinfo {year} {2009})}\BibitemShut
  {NoStop}%
\bibitem [{\citenamefont {Osterloh}\ \emph {et~al.}(2002)\citenamefont
  {Osterloh}, \citenamefont {Amico}, \citenamefont {Falci},\ and\ \citenamefont
  {Fazio}}]{osterloh2002scaling}%
  \BibitemOpen
  \bibfield  {author} {\bibinfo {author} {\bibfnamefont {A.}~\bibnamefont
  {Osterloh}}, \bibinfo {author} {\bibfnamefont {L.}~\bibnamefont {Amico}},
  \bibinfo {author} {\bibfnamefont {G.}~\bibnamefont {Falci}}, \ and\ \bibinfo
  {author} {\bibfnamefont {R.}~\bibnamefont {Fazio}},\ }\href {\doibase
  10.1038/416608a} {\bibfield  {journal} {\bibinfo  {journal} {Nature}\
  }\textbf {\bibinfo {volume} {416}},\ \bibinfo {pages} {608} (\bibinfo {year}
  {2002})}\BibitemShut {NoStop}%
\bibitem [{\citenamefont {Zurek}(2000)}]{zurek2000einselection}%
  \BibitemOpen
  \bibfield  {author} {\bibinfo {author} {\bibfnamefont {W.}~\bibnamefont
  {Zurek}},\ }\href {\doibase
  10.1002/1521-3889(200011)9:11/12<855::AID-ANDP855>3.0.CO;2-K} {\bibfield
  {journal} {\bibinfo  {journal} {Ann. Phys.}\ }\textbf {\bibinfo {volume}
  {9}},\ \bibinfo {pages} {855} (\bibinfo {year} {2000})}\BibitemShut {NoStop}%
\bibitem [{\citenamefont {Ollivier}\ and\ \citenamefont
  {Zurek}(2001)}]{ollivier2001quantum}%
  \BibitemOpen
  \bibfield  {author} {\bibinfo {author} {\bibfnamefont {H.}~\bibnamefont
  {Ollivier}}\ and\ \bibinfo {author} {\bibfnamefont {W.~H.}\ \bibnamefont
  {Zurek}},\ }\href {\doibase 10.1103/PhysRevLett.88.017901} {\bibfield
  {journal} {\bibinfo  {journal} {Phys. Rev. Lett.}\ }\textbf {\bibinfo
  {volume} {88}},\ \bibinfo {pages} {017901} (\bibinfo {year}
  {2001})}\BibitemShut {NoStop}%
\bibitem [{\citenamefont {Osborne}\ and\ \citenamefont
  {Nielsen}(2002)}]{osborne2002entanglement}%
  \BibitemOpen
  \bibfield  {author} {\bibinfo {author} {\bibfnamefont {T.~J.}\ \bibnamefont
  {Osborne}}\ and\ \bibinfo {author} {\bibfnamefont {M.~A.}\ \bibnamefont
  {Nielsen}},\ }\href {\doibase 10.1103/PhysRevA.66.032110} {\bibfield
  {journal} {\bibinfo  {journal} {Phys. Rev. A}\ }\textbf {\bibinfo {volume}
  {66}},\ \bibinfo {pages} {032110} (\bibinfo {year} {2002})}\BibitemShut
  {NoStop}%
\bibitem [{\citenamefont {Pirandola}(2014)}]{pirandola2013quantum}%
  \BibitemOpen
  \bibfield  {author} {\bibinfo {author} {\bibfnamefont {S.}~\bibnamefont
  {Pirandola}},\ }\href {\doibase 10.1038/srep06956} {\bibfield  {journal}
  {\bibinfo  {journal} {Sci. Rep.}\ }\textbf {\bibinfo {volume} {4}},\
  (\bibinfo {year} {2014})}\BibitemShut {NoStop}%
\bibitem [{\citenamefont {Streltsov}\ \emph {et~al.}(2012)\citenamefont
  {Streltsov}, \citenamefont {Kampermann},\ and\ \citenamefont
  {Bru{\ss}}}]{streltsov2012quantum}%
  \BibitemOpen
  \bibfield  {author} {\bibinfo {author} {\bibfnamefont {A.}~\bibnamefont
  {Streltsov}}, \bibinfo {author} {\bibfnamefont {H.}~\bibnamefont
  {Kampermann}}, \ and\ \bibinfo {author} {\bibfnamefont {D.}~\bibnamefont
  {Bru{\ss}}},\ }\href {\doibase 10.1103/PhysRevLett.108.250501} {\bibfield
  {journal} {\bibinfo  {journal} {Phys. Rev. Lett.}\ }\textbf {\bibinfo
  {volume} {108}},\ \bibinfo {pages} {250501} (\bibinfo {year}
  {2012})}\BibitemShut {NoStop}%
\bibitem [{\citenamefont {Baumgratz}\ \emph {et~al.}(2014)\citenamefont
  {Baumgratz}, \citenamefont {Cramer},\ and\ \citenamefont
  {Plenio}}]{baumgratz2014quantifying}%
  \BibitemOpen
  \bibfield  {author} {\bibinfo {author} {\bibfnamefont {T.}~\bibnamefont
  {Baumgratz}}, \bibinfo {author} {\bibfnamefont {M.}~\bibnamefont {Cramer}}, \
  and\ \bibinfo {author} {\bibfnamefont {M.~B.}\ \bibnamefont {Plenio}},\
  }\href {\doibase 10.1103/PhysRevLett.113.140401} {\bibfield  {journal}
  {\bibinfo  {journal} {Phys. Rev. Lett.}\ }\textbf {\bibinfo {volume} {113}},\
  \bibinfo {pages} {140401} (\bibinfo {year} {2014})}\BibitemShut {NoStop}%
\bibitem [{\citenamefont {Winter}\ and\ \citenamefont
  {Yang}(2016)}]{winter2016operational}%
  \BibitemOpen
  \bibfield  {author} {\bibinfo {author} {\bibfnamefont {A.}~\bibnamefont
  {Winter}}\ and\ \bibinfo {author} {\bibfnamefont {D.}~\bibnamefont {Yang}},\
  }\href {\doibase 10.1103/PhysRevLett.116.120404} {\bibfield  {journal}
  {\bibinfo  {journal} {Phys. Rev. Lett.}\ }\textbf {\bibinfo {volume} {116}},\
  \bibinfo {pages} {120404} (\bibinfo {year} {2016})}\BibitemShut {NoStop}%
\bibitem [{\citenamefont {Chitambar}\ \emph {et~al.}(2016)\citenamefont
  {Chitambar}, \citenamefont {Streltsov}, \citenamefont {Rana}, \citenamefont
  {Bera}, \citenamefont {Adesso},\ and\ \citenamefont
  {Lewenstein}}]{chitambar2016assisted}%
  \BibitemOpen
  \bibfield  {author} {\bibinfo {author} {\bibfnamefont {E.}~\bibnamefont
  {Chitambar}}, \bibinfo {author} {\bibfnamefont {A.}~\bibnamefont
  {Streltsov}}, \bibinfo {author} {\bibfnamefont {S.}~\bibnamefont {Rana}},
  \bibinfo {author} {\bibfnamefont {M.}~\bibnamefont {Bera}}, \bibinfo {author}
  {\bibfnamefont {G.}~\bibnamefont {Adesso}}, \ and\ \bibinfo {author}
  {\bibfnamefont {M.}~\bibnamefont {Lewenstein}},\ }\href {\doibase
  10.1103/PhysRevLett.116.070402} {\bibfield  {journal} {\bibinfo  {journal}
  {Phys. Rev. Lett.}\ }\textbf {\bibinfo {volume} {116}},\ \bibinfo {pages}
  {070402} (\bibinfo {year} {2016})}\BibitemShut {NoStop}%
\bibitem [{\citenamefont {Hillery}(2016)}]{hillery2016coherence}%
  \BibitemOpen
  \bibfield  {author} {\bibinfo {author} {\bibfnamefont {M.}~\bibnamefont
  {Hillery}},\ }\href {\doibase 10.1103/PhysRevA.93.012111} {\bibfield
  {journal} {\bibinfo  {journal} {Phys. Rev. A}\ }\textbf {\bibinfo {volume}
  {93}},\ \bibinfo {pages} {012111} (\bibinfo {year} {2016})}\BibitemShut
  {NoStop}%
\bibitem [{\citenamefont {Shi}\ \emph {et~al.}(2017)\citenamefont {Shi},
  \citenamefont {Liu}, \citenamefont {Wang}, \citenamefont {Yang},
  \citenamefont {Yang},\ and\ \citenamefont {Fan}}]{shi2016coherence}%
  \BibitemOpen
  \bibfield  {author} {\bibinfo {author} {\bibfnamefont {H.~L.}\ \bibnamefont
  {Shi}}, \bibinfo {author} {\bibfnamefont {S.~Y.}\ \bibnamefont {Liu}},
  \bibinfo {author} {\bibfnamefont {X.~H.}\ \bibnamefont {Wang}}, \bibinfo
  {author} {\bibfnamefont {W.~L.}\ \bibnamefont {Yang}}, \bibinfo {author}
  {\bibfnamefont {Z.~Y.}\ \bibnamefont {Yang}}, \ and\ \bibinfo {author}
  {\bibfnamefont {H.}~\bibnamefont {Fan}},\ }\href {\doibase
  10.1103/PhysRevA.95.032307} {\bibfield  {journal} {\bibinfo  {journal} {Phys.
  Rev. A}\ }\textbf {\bibinfo {volume} {95}},\ \bibinfo {pages} {032307}
  (\bibinfo {year} {2017})}\BibitemShut {NoStop}%
\bibitem [{\citenamefont {Matera}\ \emph {et~al.}(2016)\citenamefont {Matera},
  \citenamefont {Egloff}, \citenamefont {Killoran},\ and\ \citenamefont
  {Plenio}}]{matera2016coherent}%
  \BibitemOpen
  \bibfield  {author} {\bibinfo {author} {\bibfnamefont {J.~M.}\ \bibnamefont
  {Matera}}, \bibinfo {author} {\bibfnamefont {D.}~\bibnamefont {Egloff}},
  \bibinfo {author} {\bibfnamefont {N.}~\bibnamefont {Killoran}}, \ and\
  \bibinfo {author} {\bibfnamefont {M.~B.}\ \bibnamefont {Plenio}},\ }\href
  {\doibase 10.1088/2058-9565/1/1/01LT01} {\bibfield  {journal} {\bibinfo
  {journal} {Quantum Sci. Technol.}\ }\textbf {\bibinfo {volume} {1}},\
  \bibinfo {pages} {01LT01} (\bibinfo {year} {2016})}\BibitemShut {NoStop}%
\bibitem [{\citenamefont {Andrade}\ and\ \citenamefont
  {da~Luz}(2009)}]{andrade2009equivalence}%
  \BibitemOpen
  \bibfield  {author} {\bibinfo {author} {\bibfnamefont {F.~M.}\ \bibnamefont
  {Andrade}}\ and\ \bibinfo {author} {\bibfnamefont {M.~G.~E.}\ \bibnamefont
  {da~Luz}},\ }\href {\doibase 10.1103/PhysRevA.80.052301} {\bibfield
  {journal} {\bibinfo  {journal} {Phys. Rev. A}\ }\textbf {\bibinfo {volume}
  {80}},\ \bibinfo {pages} {052301} (\bibinfo {year} {2009})}\BibitemShut
  {NoStop}%
\bibitem [{\citenamefont {Brassard}\ \emph {et~al.}(2002)\citenamefont
  {Brassard}, \citenamefont {Hoyer}, \citenamefont {Mosca},\ and\ \citenamefont
  {Tapp}}]{brassard2002quantum}%
  \BibitemOpen
  \bibfield  {author} {\bibinfo {author} {\bibfnamefont {G.}~\bibnamefont
  {Brassard}}, \bibinfo {author} {\bibfnamefont {P.}~\bibnamefont {Hoyer}},
  \bibinfo {author} {\bibfnamefont {M.}~\bibnamefont {Mosca}}, \ and\ \bibinfo
  {author} {\bibfnamefont {A.}~\bibnamefont {Tapp}},\ }in\ \href@noop {} {\emph
  {\bibinfo {booktitle} {Quantum Computation and Information}}},\ \bibinfo
  {editor} {edited by\ \bibinfo {editor} {\bibfnamefont {S.~J.}\ \bibnamefont
  {Lomonaco}}\ and\ \bibinfo {editor} {\bibfnamefont {H.~E.}\ \bibnamefont
  {Brandt}}}\ (\bibinfo  {publisher} {AMS},\ \bibinfo {address} {New York},\
  \bibinfo {year} {2002})\BibitemShut {NoStop}%
\end{thebibliography}%
\end{document}